\documentclass[a4paper,11pt]{article}

\pdfoutput=1\relax                   
\usepackage{mystyle}

\usepackage[11pt]{moresize}

\graphicspath{{./}{./figures/}} 

\title{}

\newcommand{\longTitle}{Leveraging Peer Feedback to Improve Visualization Education}
\newcommand{\shortTitle}{Peer Feedback in Visualization Education}

\hypersetup{
    pdftitle={\longTitle},    
    pdfauthor={Paul Rosen},     
    pdfsubject={\shortTitle},   
    pdfcreator={Paul Rosen},   
    pdfproducer={Paul Rosen}, 
    pdfkeywords={visualization education}{peer review}{pedagogy}, 
}

\newcommand{\colorBorderBox}[1]{\setlength{\fboxrule}{2pt}\fcolorbox{#1}{white}{\begin{minipage}[m]{3pt}\color{white}x\vspace{-6pt}\end{minipage}}\setlength{\fboxrule}{1pt}}
\newcommand{\studentQuote}[1]{\vspace{12pt}\noindent\hspace{0.1\linewidth}\begin{minipage}{0.80\linewidth}\emph{``#1''}\end{minipage}}
\newcommand{\ttest}[3]{$t(#1)$$=$$#2$, $p$$=$$#3$}
\newcommand{\teaser}[1]{\begin{figure}[!ht]#1\end{figure}}
\renewcommand{\abstract}[1]{\noindent\hspace{0.05\linewidth}\begin{minipage}{0.90\linewidth}\textbf{Abstract.} #1\end{minipage}}
\newcommand{\ColorCode}{\setlength{\fboxrule}{0.5pt}\setlength\fboxsep{3pt}
Color-code:
	\fcolorbox{black}{color2017Survey}{\color{color2017Survey}\tiny a}~2017~/ 
    \fcolorbox{black}{color2018Survey}{\color{color2018Survey}\tiny a}~2018~/ 
    \fcolorbox{black}{color2019Survey}{\color{color2019Survey}\tiny a}~2019}

\newcommand{\refLOne}{{[L1]}}   
\newcommand{\refLTwo}{{[L2]}}   
\newcommand{\refLThree}{{[L3]}} 

\begin{document}

\definecolor{color2017Survey}{HTML}{b3cde3}
\definecolor{color2018Survey}{HTML}{ccebc5}
\definecolor{color2019Survey}{HTML}{FFB1B1}

\definecolor{colorFamil}{HTML}{548235}
\definecolor{colorFound}{HTML}{7030A0}
\definecolor{colorSkill}{HTML}{C55A11}
\definecolor{colorSyste}{HTML}{FFC000}
\definecolor{colorPeerR}{HTML}{7F7F7F}
\definecolor{colorProje}{HTML}{F2F2F2}

\begin{center}{\bf \Large \longTitle}\footnote{Cite as: Z. Beasley, A. Friedman, L. Pieg and P. Rosen, "Leveraging Peer Feedback to Improve Visualization Education," 2020 IEEE Pacific Visualization Symposium (PacificVis), Tianjin, China, 2020, pp. 146-155, DOI: 10.1109/PacificVis48177.2020.1261.}\footnote{\textcopyright\ 2020 IEEE. Personal use of this material is permitted. Permission from IEEE must be
obtained for all other uses, in any current or future media, including
reprinting/republishing this material for advertising or promotional purposes, creating new
collective works, for resale or redistribution to servers or lists, or reuse of any copyrighted
component of this work in other works.} \end{center}

\begin{minipage}[t]{0.24\linewidth}
\centering
Zachariah Beasley \\ 
\footnotesize
zjb@mail.usf.edu\\
University of South Florida
\end{minipage}
\begin{minipage}[t]{0.24\linewidth}
\centering
Alon Friedman \\ 
\footnotesize
alonfriedman@usf.edu \\
University of South Florida
\end{minipage}
\begin{minipage}[t]{0.24\linewidth}
\centering
Les Piegl \\
\footnotesize
lespiegl@mail.usf.edu \\
University of South Florida
\end{minipage}
\begin{minipage}[t]{0.24\linewidth}
\centering
Paul Rosen \\
\footnotesize
prosen@usf.edu \\
University of South Florida
\end{minipage}

\teaser{
    \includegraphics[width=1\linewidth]{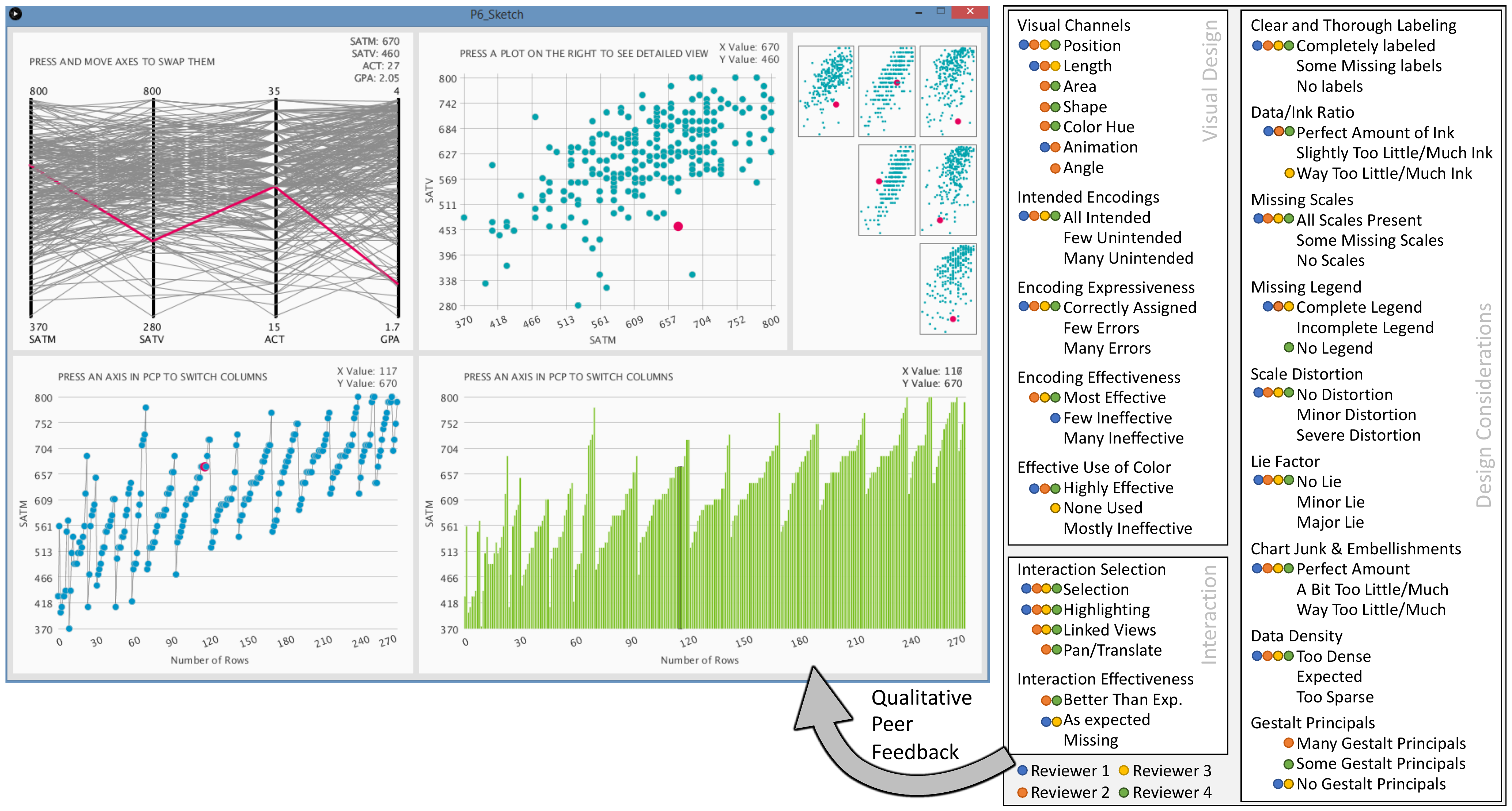}
    \begin{minipage}[t]{0.51\linewidth}
        \vspace{-48pt}
        \caption{Left: Example submission for Project 6 ``Building a Dashboard'', described in \autoref{sec.course}. Right: Visualization of the qualitative feedback received from 4 peers to the project submission based upon the rubric described in \autoref{sec.rubric}.}
	\label{fig:teaser}
    \end{minipage}
}

\newcommand{\placeImageLabelA}[3]{\put(-18,45){
  		\begin{minipage}[t][0pt][t]{0pt}
			\tiny 
			\mbox{#1} 
		\end{minipage}
		}}

\newcommand{\placeImageLabelB}[3]{\put(-120,75){
  		\begin{minipage}[t][0pt][t]{0pt}
			\small 
			\mbox{#1}
		\end{minipage}
		}}

\newcommand{\placeImageLabelC}[3]{\put(-110,75){
  		\begin{minipage}[t][0pt][t]{0pt}
			\small 
			\mbox{#1}
		\end{minipage}
		}}

\newcommand{\placeImageLabelF}[3]{\put(-120,75){
  		\begin{minipage}[t][0pt][t]{0pt}
			\small 
			\mbox{#1}
		\end{minipage}
		}}

\abstract{
Peer review is a widely utilized pedagogical feedback mechanism for engaging students, which has been shown to improve educational outcomes. However, we find limited discussion and empirical measurement of peer review in visualization coursework. In addition to engagement, peer review provides direct and diverse feedback and reinforces recently-learned course concepts through critical evaluation of others' work. In this paper, we discuss the construction and application of peer review in a computer science visualization course, including: projects that reuse code and visualizations in a feedback-guided, continual improvement process and a peer review rubric to reinforce key course concepts. To measure the effectiveness of the approach, we evaluate student projects, peer review text, and a post-course questionnaire from 3 semesters of mixed undergraduate and graduate courses. The results indicate that course concepts are reinforced with peer review---82\% reported learning more because of peer review, and 75\% of students recommended continuing it. Finally, we provide a road-map for adapting peer review to other visualization courses to produce more highly engaged students.
}

\section{Introduction}
\label{sec.intro}

Rushmeier et al.~\cite{rushmeier2007revisiting} defined visualization education as a work in progress. Nevertheless, the subject can be broadly split into 2 categories. The first is the proper construction of visualizations---using the right algorithms and visualization principles in creating visualizations, which tends to be the primary focus of visualization courses, and student comprehension of concepts, techniques, and algorithms can be objectively measured. The second category is focused on the subjective evaluation of the quality and accuracy of visualizations. Subjective evaluation is not only important for the instructor's assessment of students but for students to \textit{develop the ability to evaluate others' visualizations critically}. These skills are commonly taught through informal methods, such as group or whole-class discussions that can leave students' skills underdeveloped \cite{santos2018heuristic}.

Furthermore, with Gen Z learners (individuals born between the mid-1990s and mid-2000s), educational preferences have shifted significantly. They prefer instant feedback, are increasingly collaborative, and are active-learners who prefer project-based coursework~\cite{Renfro2012}.  Meanwhile, with a large number of students enrolled in visualization courses, it is difficult to provide students the timely, subjective feedback they need to improve the quality of their work~\cite{sorby2000development}.

Peer review is a highly-engaging feedback mechanism \cite{kearney2013improving, weaver2012peer}, often used in liberal arts courses~\cite{beaufort2008college, Nowacek2011Agents, wardle2012addressing, moxley2013big}, human-computer interaction (HCI) courses~\cite{shamim2015evaluation, eppler2006comparison, Novak2006Theorigins}, code review~\cite{bacchelli2013expectations}, and scholarly publication~\cite{ware2011peer}, and it is ideal for addressing these challenges. Instead of relying solely on instructors for feedback, peers collaborate to provide diverse multi-sourced feedback to one another with a relatively quick turnaround. In addition, the evaluation process itself gives students an opportunity to reinforce recently-learned course concepts by critically evaluating others' work. Finally, for instructors, it is a scalable approach. Since students provide feedback to one another, adding students to a course only adds to nominal administrative efforts.

Despite these well-known advantages of peer review, we found little evidence of peer review in visualization courses. In a survey of 100 information visualization faculty, we found 18 publicly available course syllabi, only 1 of which mentioned peer review. \textit{In order to address this gap, we discuss the construction and evaluation of our own peer review-oriented computer science visualization course, in order to encourage the visualization community to initiate discussions around and to eventually adopt this pedagogical methodology into their classrooms.}

\vspace{5pt}
\noindent
The contributions of this paper are:
\vspace{-5pt}
\begin{enumerate}[noitemsep,itemsep=5pt]
    \item We describe a peer review-oriented visualization course with projects designed to reuse code and visualizations in a feedback-guided continual improvement process; 
    \item We discuss and evaluate a peer review rubric to support the continual improvement process while reinforcing key concepts of the course; 
    \item We evaluate whether peer review reinforces course material, whether students engage in and enjoy peer review, and what aspect of peer review is most beneficial to students; and 
    \item We discuss various ways to integrate peer review into existing visualization courses.
\end{enumerate}

\section{Background}

Peer review, the evaluation of work by individuals of similar competence to the producer(s) of the work~\cite{spier2002history}, has been successfully~utilized in many professional practices, e.g., code review~\cite{bacchelli2013expectations}, evaluation of scholarly and scientific works~\cite{ware2011peer}, etc. In code review, it serves as a software quality assessment activity. In science, peer review is the method by which papers are published, academic promotions secured, and Nobel prizes won. Notably, in science, the practice has been criticized as repetition based on faith rather than fact~\cite{biagioli2002book}.

\subsection{Educational Values of Peer Review}
\label{sec:background:edu}

\paragraph{Seeing Others' Work} 
In engineering education, the use of peer review in the classroom remains an open area of research~\cite{beasley2018ten}, despite the fact that it is so essential to the design and implementation of engineering systems. For example, Garousi~\cite{garousi2010applying} constructed a framework using a ``Goal-Question-Metric'' (GQM) approach and reported that students gained knowledge in preparing software development documentation and writing code and that \textit{students find peer review useful primarily since they can learn from their peers' work}. He also suggested that students might benefit from reviewing work from prior semesters before embarking on their projects.

\paragraph{Formative vs.\ Summative Value} 
Many educators have employed both \textit{formative} reviews to stimulate learning and \textit{summative} reviews for assessment purposes (i.e., providing a grade) in their courses. In liberal arts, multiple researchers have explored whether the ability to write well is a skill that can be mastered in one context and simply transferred to another~\cite{beaufort2008college, Nowacek2011Agents, wardle2012addressing}. Moxley~\cite{moxley2013big} opposed this, stating that grand claims about student ability based on a handful of rubric scores are not necessarily predictive of their classroom performance. This concern applies when considering peer review for \textit{assessment} in visualization courses, though it does not question the \textit{formative} benefits of peer review.

\paragraph{Active-Learning} 
Naps et al.~\cite{Naps2002visualengagement} suggested that no matter how well a visualization is designed, there is little educational value unless it engages students in an active-learning classroom activity. Perhaps most notably, peer review is a highly-engaging, active-learning educational mechanism~\cite{kearney2013improving, weaver2012peer}, and active-learning has been shown to significantly improve retention, comprehension, and overall educational outcomes~\cite{doi:10.1152/advan.00053.2006}.

\subsection{Visualization Education}

There is a wide breadth for the ``style'' of visualization courses that usually match the focus of the visualization sub-community  (i.e., VAST, InfoVis, SciVis), which the instructor primarily subscribes to. For a broad survey of visualization course styles, see~\cite{kerren2008teaching, owen2013visualization}. Nevertheless, much of the recent innovation in visualization education has focused on developing problem- or design-oriented courses that apply a senior project style course design to address a visualization problem over an entire semester~\cite{he2017v, whiting2009vast, scholtz2014evaluation, rohrdantz2014augmenting}. In some cases, these include the additional challenge of being interdisciplinary~\cite{domik2009my, elmqvist2012leveraging, domik2016data}.

\paragraph{Course Topics}
Proposals for formal visualization education have focused mainly on the \textit{proper construction of visualizations and other technical aspects}, not the critical evaluation of visualizations. Hanrahan and Ma~\cite{ma2005teaching} proposed the implementation of core topics to visualization education that included: data and image models; perception and cognition; interaction; space; color; shape and lines; texture; interior structure with volumetric techniques; and narrative with animation. Gilbert~\cite{gilbert2008visualization} further described 5 levels of competency---representation as depiction; converting early symbolic skills; the syntactic use of formal representation; semantic use of formal representations; and reflective, rhetorical use of presentation---that give insight to the depth of topical understanding in visualization. More recently, Munzner's \textit{Visualization Analysis \& Design}~\cite{munzner2014visualization} takes visualization concepts and splits them into what-why-how questions that she maps into data abstraction, task abstraction, and visual encoding and interaction technique taxonomies. Our course follows this text.

\paragraph{Visualization Design}
A design methodology is critical in activities from software development to visualization design. On the software design side, Agile-style iterative design~\cite{beck2001manifesto} and the Model-View-Controller design pattern help to develop software engineering skills. One of the more popular methodologies to bridge the gap between software and visualization design, Munzner's Nested Model~\cite{munzner2009nested} adapts Agile principals to the specific needs of visualization. The Design Activity Framework further connects the Nested Model to visualization design~\cite{mckenna2014design} with worksheets that guide the would-be designer~\cite{mckenna2017worksheets}. These approaches are invaluable for teaching composition and design in a visualization course. 

A different approach to design is heuristics, ``an approach, a strategy, or a trick that experience has shown may help in constructing a useful model''~\cite{Powell1995heuristicmodel}. Santos et al.\ discussed heuristic evaluation as a usability inspection method that has been adapted to evaluate visualizations~\cite{santos2018heuristic}. Various researchers have examined student performances in their visualization classes based on heuristic evaluation~\cite{hearst2016evaluating, wall2018heuristic}, and the methods provide a potential alternative design evaluation methodology that fits within the context of peer review.

\paragraph{Peer Review in Visualization}
Critiquing has long been acknowledged as a critical part of the visualization design process~\cite{kosara2007visualization}, which has, by-and-large, been taught through informal methods. \textit{We have found little evidence that the community has broadly considered peer review as an effective pedagogical method}. We surveyed  visualization course syllabi from the websites of 100 professors in the IEEE InfoVis 2019 program committee, IEEE VIS 2019 conference committees, and 2016 and 2017 Pedagogy of Data Visualization IEEE VIS Workshop contributors. Of the 18 syllabi openly available, only 1 mentioned peer review of another student/group's work\footnote{We do not include ``peer evaluation" of one’s own group as peer review.}. Four syllabi suggested at least 1 project built upon the code or requirements of a previous project; however, none of these utilized peer review.

Although we understand that peer review may be utilized in more visualization courses than we identified in our survey, the results suggest a need for expanded evaluation and discussion of the benefits of peer review in the visualization classroom. The only peer review evaluation that we are aware of is our work from the 2017 Pedagogy of Data Visualization IEEE VIS workshop, where we presented a rubric for peer review for visualization courses~\cite{Friedman2017}. The rubric included 5 main categories of mix-and-match style evaluation criteria that can be customized to a wide variety of contexts. This work is built around that rubric and the formalization of critique within a visualization course through peer review.

\section{Course Overview}
\label{sec.course}

With the extensive usage of peer review in other disciplines, we endeavored to build a visualization course with peer review as a core component. Our course, titled ``Data Visualization'', was taught Spring semester of 2017, 2018, and 2019 and was a co-listed undergraduate and graduate course. The course was hosted in the Computer Science department, within the College of Engineering at the University of South Florida, and the educational emphasis of the course was primarily good visualization practice, though a strong emphasis was placed on software design as well.

\subsection{Course Content}

The content of the course was primarily taught following Munzner's \textit{Visualization Analysis \& Design}~\cite{munzner2014visualization} and the Nested Model~\cite{munzner2009nested} with additional outside visualization content (e.g., Vis Lies\footnote{Vis Lies: \url{http://www.vislies.org}}, New York Times Graphics\footnote{New York Times: \url{https://www.nytimes.com}}, etc.) and research papers added throughout the semester. We decided to pursue structured assessment (i.e., well-defined projects), leaving Problem Characterization aspects of the Nested Model to in-class presentation, activities, and discussions. The remaining levels, Data Abstraction, Encoding and Interaction, and Algorithm Design, were covered in class and emphasized through the projects. The presentation methods were primarily lecturing, research paper presentations by students, and discussions (e.g., small group and whole class critiques).

The course learning objectives (see syllabus in supplementary materials) were that students would demonstrate the ability to:

\begin{enumerate}[noitemsep,itemsep=5pt]
    \item[\refLOne] Build effective visualizations by evaluating a provided data and user requirements and programming an interface to match those requirements.
    \item[\refLTwo] Associate visualizations with the foundational components, e.g., data abstractions and visual encodings, that go into their construction.
    \item[\refLThree] Critique the effectiveness of interactive visualizations with respect to task selection, visual encoding choices, and interaction design and implementation.    
\end{enumerate}

To satisfy \textbf{\refLOne} and \textbf{\refLTwo}, the course consisted of 8 projects (see \autoref{sec.course}), 1 in Tableau\footnote{Tableau: \url{https://www.tableau.com/}} and 7 in Processing\footnote{Processing: \url{https://processing.org/}}, totaling $50\%$ of their final grade. Projects were due every 10-14 days, except for Project~6 that allowed approximately 30 days since it was assigned over Spring Break. To satisfy \textbf{\refLThree}, peer review of projects was integrated into the course. Upon the completion of each project, students were asked to provide reviews of 3 randomly selected peers' work using a provided rubric within 5-7 days (see \autoref{sec.rubric}). The peer feedback served as the primary form of qualitative feedback students received. In addition, the Spring 2018 version of the class included self-review after the completion of peer review. Both peer review and self-review were for small amounts of (spot checked) \textit{completion-based} credit (approximately $10\%$ of the final course grade).  The instructor and teaching assistants still completed project grading. Grades were assigned based primarily upon objective requirements, as well as some subjective judgment.

\subsection{Evaluation Methodology}
\label{sec.course.eval}

There are significant subjective measurements necessary to evaluate the effectiveness of a visualization course~\cite{viegas2015design}. Visualization education research also lacks a standard rubric for measuring engagement. In a review of 5 major visualization venues (InfoVis, SciVis, VAST, EuroVis, and Pacific Vis), we found no form of empirical evaluation of students' work or engagement in the classroom.  Thus, we have conducted a qualitative evaluation using questionnaires to measure \textit{perceived} improvement and engagement. We supplement and reinforce the analysis by combining the questionnaire data with both qualitative analysis using representative examples of student work and quantitative analysis through natural language processing.

A useful reference for our non-experimental evaluation methodology can be found in the field of writing composition. For example, Mulder et al.\ used similar qualitative and quantitative methodologies, based on content analysis from peer review comments and student questionnaires~\cite{Mulder2014casestudy}. Furthermore, analyzing the number and variety of words used in written comments has been used for measuring both learning outcomes~\cite{McGourty19989Stuentpeerreview} and student engagement~\cite{van2010effective}.

\subsection{Data Collection}
\label{sec.course.data}

The student-produced data falls into the following categories:

\paragraph{Student Visualizations} Project submissions were gathered from 8 projects each year (2017-2019). We manually reviewed projects to find those which, in combination with peer comments, exemplify the value of peer review. We have included 1 such sample project in the supplemental materials\footnote{Due to the sheer number of projects (over 1000), we were unable to anonymize all projects for inclusion in the supplemental materials.}.

\begin{figure}[!b]
    \centering
    \includegraphics[width=0.9\linewidth]{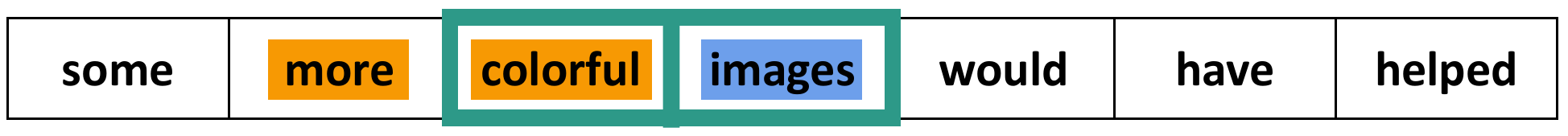}
    \caption{Aspect Extractor: noun in close proximity to an adjective.}
    \label{fig:ae}
\end{figure}

\paragraph{Student Peer Review and Self-Review Comments} 
Student peer review comments were collected on 8 projects from 2017 and 7~projects from 2018 and 2019. Student self-review comments were collected on 7 projects from 2018 only\footnote{Self-review was dropped in 2019 due to concerns about workload.}. We analyzed them with a dictionary-based natural language processing algorithm~\cite{beasley2019designing} for matching positive and negative keywords to produce numerical feedback including overall sentiment (positive or negative) of the text, counts of parts of speech (i.e., noun, adjective, adverb), the average length of comments, etc. The algorithm includes an \textit{aspect extractor} (similar to one developed by Google towards analyzing reviews of local services, such as restaurants and hotels~\cite{blair2008building}) that scans text in a sliding window and produces a list of important aspects (i.e., nouns), which are in close proximity to sentiment words (i.e., adjectives). The idea is that these aspects are frequently commented upon words associated with either positive or negative sentiment, indicating their importance to reviewers. \autoref{fig:ae} shows an example adjective (``colorful") and noun (``images") match within the sliding window, highlighted in teal. While ``colorful" has inferred positive sentiment, it is modified by a comparative adjective, ``more", which reverses its sentiment to negative. It is notable that this algorithm was developed for peer review in engineering courses, but not explicitly tuned for visualization courses or the particular rubric utilized. Thus, the algorithm does not ``look" for visualization keywords, only for general sentiment-producing words.

\begin{figure}[!t]
    \centering

    {\begin{minipage}[b]{0.185\linewidth}
        \centering
        \includegraphics[height=1.25in]{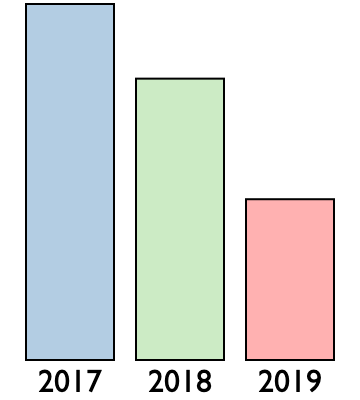} \footnotesize n=139
        \vspace{-5pt}
        \subcaption[scriptsize]{}
    \end{minipage}}
    \hfill
    {\begin{minipage}[b]{0.12\linewidth}
        \centering
        \includegraphics[height=1.25in]{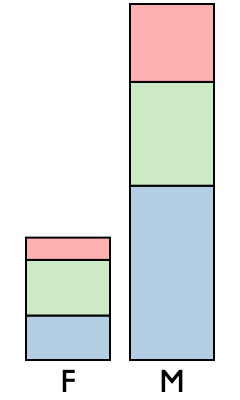} \footnotesize n=129
        \vspace{-5pt}
        \subcaption[scriptsize]{}
    \end{minipage}}    
    \hfill
    {\begin{minipage}[b]{0.12\linewidth}
        \centering
        \includegraphics[height=1.25in]{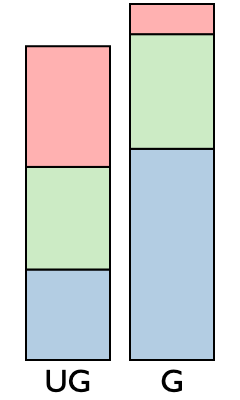} \footnotesize n=111
        \vspace{-5pt}
        \subcaption[scriptsize]{}
    \end{minipage}} 
    \hfill
    {\begin{minipage}[b]{0.185\linewidth}
        \centering
        \includegraphics[trim=245pt 0 0 0, clip, height=1.25in]{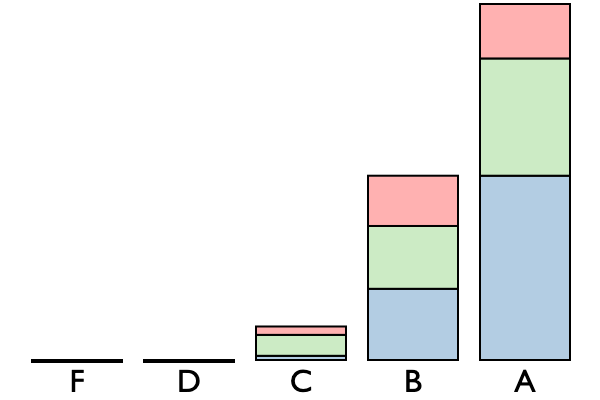} \footnotesize n=137
        \vspace{-5pt}
        \subcaption[scriptsize]{}
    \end{minipage}} 
    \hfill
    {\begin{minipage}[b]{0.245\linewidth}
        \centering
        \includegraphics[height=1.25in]{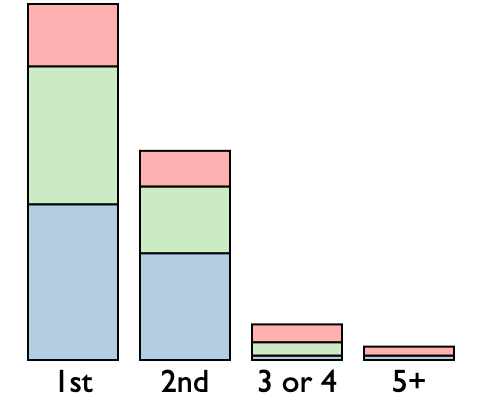} \footnotesize n=138
        \vspace{-5pt}
        \subcaption[scriptsize]{}
    \end{minipage}} 
    
    \caption{Histograms of questionnaire respondent demographics, including (a) students per year, (b) gender, (c) undergraduate vs.\ graduate, (d)~anticipated course grade, (e)~times using peer review in a CS course. \ColorCode}
    \label{fig:demographics}
\end{figure}

\paragraph{Post-Course Questionnaire} 
Each year (2017-2019), students were asked to complete an optional 20-question post-course questionnaire, given at the conclusion of their final exam and before receiving their final grade. Students were offered a small amount of extra credit to anonymously answer questions ranging from demographic information (see \autoref{fig:demographics}) and anticipated final grade to helpfulness of review comments and open-ended suggestions for improving the peer review process. Numerical answers were all placed on a 5-point Likert scale. The overall participation rate was $98\%$. We performed quantitative analysis on numerical answers and manually reviewed written feedback for qualitative results. The questionnaires and student responses are included in our supplementary materials.

\section{Project Design}

When designing projects, our interest was to see students gain visualization skills (i.e., \textbf{\refLTwo}) by demonstrating proficiency in using software engineering problem-solving techniques (i.e., \textbf{\refLOne})~\cite{cleveland1993visualizing}. 
To support this dual goal of visualization practice and software design, the course projects were divided into 4 categories of emphasis:

\vspace{7pt}\noindent\colorBorderBox{colorFamil}\hspace{3pt}{\begin{minipage}[t]{0.925\linewidth}\underline{Familiarization and Expectation Setting}: This project familiarizes students with using standard visualization types.\end{minipage}}

\vspace{7pt}\noindent\colorBorderBox{colorFound}\hspace{3pt}{\begin{minipage}[t]{0.925\linewidth}\underline{Visualization Foundation Building}: These projects emphasize the implementation of basic mechanics of visualization, such as data abstraction, visual encoding, and interaction.\end{minipage}}

\vspace{7pt}\noindent\colorBorderBox{colorSkill}\hspace{3pt}{\begin{minipage}[t]{0.925\linewidth}\underline{Applying Skills to New Contexts}: In these projects, students apply their skills to develop more complex visualizations.\end{minipage}}

\vspace{7pt}\noindent\colorBorderBox{colorSyste}\hspace{3pt}{\begin{minipage}[t]{0.925\linewidth}\underline{Software System Engineering}: These projects use software engineering skills to build interactive multi-view interfaces.\end{minipage}}

\vspace{7pt}
The peer reviews serve as an integral part of the future projects by allowing students to use feedback to refine the work they submit (i.e., \textbf{\refLOne}). Therefore, the projects are set up to maximally build upon and reuse components from previous projects, while still challenging students with new project requirements. \autoref{fig:project_map} shows the timeline of projects over the semester. The solid black edges indicate direct reuse of components and associated feedback from previous projects. Full project descriptions are included in our supplementary material.

\subsection{Project Descriptions}

\noindent\colorBorderBox{colorFamil}\hspace{3pt}{\begin{minipage}[b]{0.85\linewidth} \textsc{Project 1: Building a Visualization Story}\vspace{-3pt}\end{minipage}} 
\vspace{3pt}

\noindent
\textit{Objective:} The goal of the project, which was assigned the first day of class, is for students to become familiar with a fully functional visualization system and to get them thinking about objective/task-based visualization design. \textit{Requirements:} Using Tableau, the students are to: import a dataset of their choice, develop 3 ``questions'' about the data, choose 3 or more visualizations to answer those questions, and finally place the results into a short narrative.

\vspace{5pt}\noindent\colorBorderBox{colorFound}\hspace{3pt}{\begin{minipage}[b]{0.85\linewidth} \textsc{Project 2: Familiarization with Processing}\vspace{-3pt}\end{minipage}} \vspace{3pt}

\noindent
\textit{Objective:} This project provides students a first glimpse at the Processing development environment by developing simple interfaces with simple data. \textit{Requirements:} The students are asked to create 3~sketches: a bar chart, a line chart, and a combined bar and line chart. Although the students have had little exposure in class, they are encouraged to use their best judgment concerning coloring, scales, labeling, tick marks, and other embellishments.

\vspace{5pt}\noindent\colorBorderBox{colorFound}\hspace{3pt}{\begin{minipage}[b]{0.75\linewidth} \textsc{Project 3: Reusing Drawing Objects}\vspace{-3pt}\end{minipage}} 
\vspace{3pt}

\noindent
\textit{Objective:} The project provides a lesson in building reusable interfaces.
\textit{Requirements:} The students are to build 2~sketches: first, a scatterplot, and second, a scatterplot matrix (reusing the scatterplot object) with a dataset of approximately 270~data items and 4 attributes. The students are encouraged to use additional visual channels, wherever possible, and again to use their best judgment concerning scales, labeling, tick marks, and other embellishments.

\vspace{5pt}\noindent\colorBorderBox{colorFound}\hspace{3pt}{\begin{minipage}[b]{0.75\linewidth} \textsc{Project 4: Adding Interaction}\vspace{-3pt}\end{minipage}} 
\vspace{3pt}

\noindent
\textit{Objective:} In this project, students are encouraged to explore options for improving their visualizations using interaction. \textit{Requirements:} Students are required to add interactions to their sketches from Projects 2 and 3. For the line chart, bar chart, and scatterplot, this is in the form of data item selection. For the scatterplot matrix, the selection provides the attributes used in a detail view scatterplot. \textit{Utilization of Feedback:} This project provides the first opportunity for students to directly apply the feedback they have received. They are actively encouraged to use the feedback from Projects 2 and 3 to improve the aesthetics and functionality of their visualizations.

\vspace{5pt}\noindent\colorBorderBox{colorSkill}\hspace{3pt}{\begin{minipage}[b]{0.85\linewidth} \textsc{Project 5: Advanced Visualization Interface}\vspace{-3pt}\end{minipage}} \vspace{3pt}

\noindent
\textit{Objective:} This project asks students to use the knowledge they have built in the previous projects to build an advanced visualization interface---parallel coordinates plot---from scratch. \textit{Requirements:} Build a parallel coordinates plot with interactions, such as axis swapping and brushing.

\vspace{5pt}\noindent\colorBorderBox{colorSyste}\hspace{3pt}{\begin{minipage}[b]{0.75\linewidth} \textsc{Project 6: Building a Dashboard}\vspace{-3pt}\end{minipage}} 
\vspace{3pt}

\noindent
\textit{Objective:} The objective of this project is to combine previous projects into a single linked-view dashboard interface. \textit{Requirements:} Create a dashboard using the bar chart, line chart, scatterplot, scatterplot matrix, and parallel coordinates plot from Projects 2-5. Provide linked-view interaction via selection. Students are encouraged to add additional interactions and to use their best judgment concerning dashboard layout.
\textit{Utilization of Feedback:} Since this project combines the work of Projects 2-5, students are encouraged to revisit their previous designs and associated feedback to improve the look and functionality of their dashboards.

\vspace{5pt}\noindent\colorBorderBox{colorSyste}\hspace{3pt}{\begin{minipage}[b]{0.75\linewidth} \textsc{Project 7: Adding Aggregation}\vspace{-3pt}\end{minipage}} 
\vspace{3pt}

\noindent
\textit{Objective:} The objective of this project is to extend the dashboard with capabilities that better support large data through statistically derived attributes. \textit{Requirements:} Extend the dashboard from Project~6 to include histograms and corrgrams (using both Pearson Correlation Coefficient and Spearman Rank Correlation Coefficient) of the data. \textit{Utilization of Feedback:} The students are encouraged to use their feedback from Project 6 (in addition to Projects 2-5) to improve the design of their dashboard.

\vspace{5pt}\noindent\colorBorderBox{colorSkill}\hspace{3pt}{\begin{minipage}[b]{0.75\linewidth} \textsc{Project 8: Drawing Graphs}\vspace{-3pt}\end{minipage}} 
\vspace{3pt}

\noindent
\textit{Objective:} This project is mostly independent of the others, enabling students to translate lessons learned to a new context. \textit{Requirements:} Visualize a graph using a force-directed layout. In addition, the students must provide the ability to interact with the layout.

\setlength{\fboxrule}{1pt}
\setlength\fboxsep{1pt}
    
\begin{figure}[!b]
    \centering
    
    \includegraphics[width=0.65\linewidth]{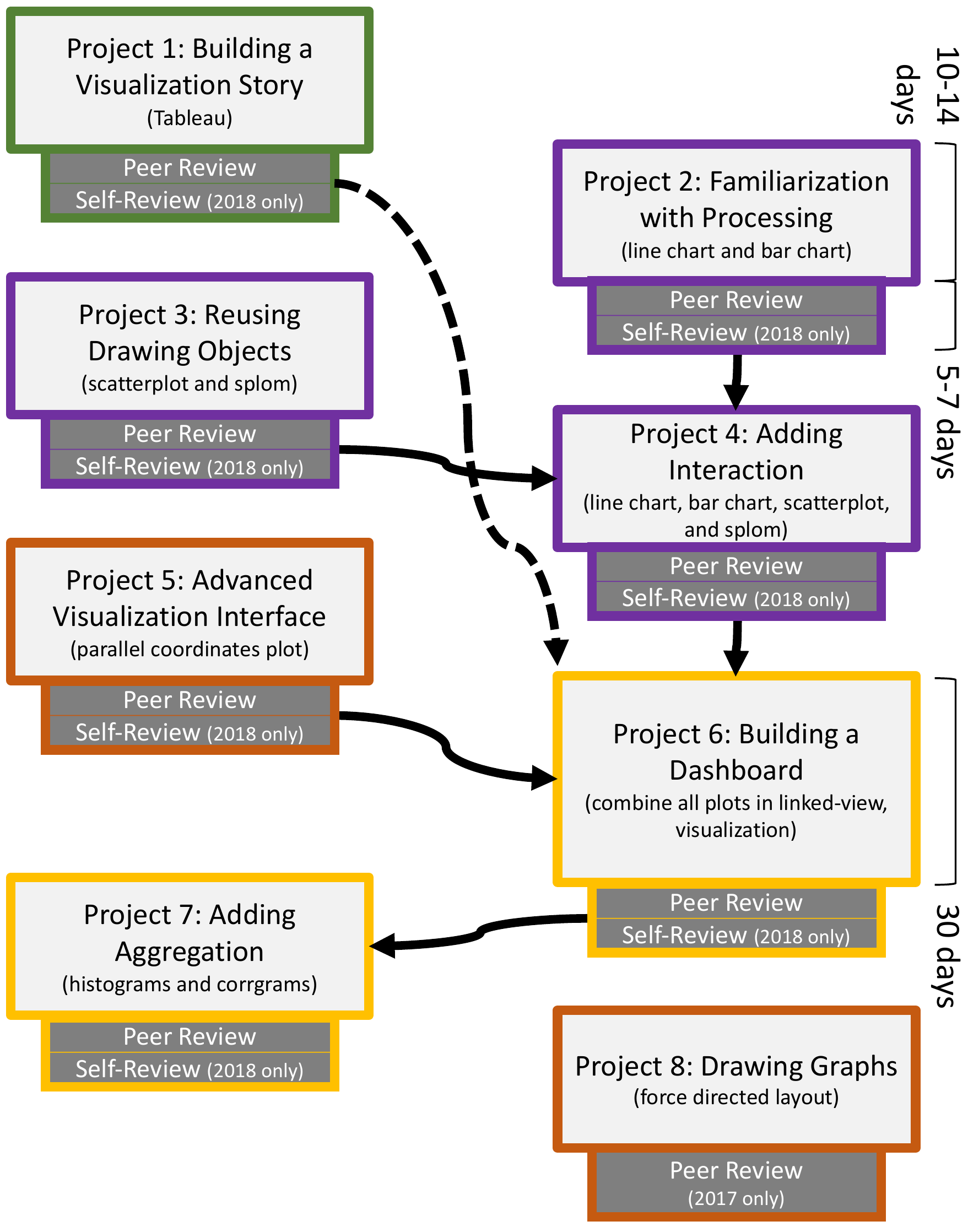}
    \caption{A timeline of course \fcolorbox{lightgray}{colorProje}{\ssmall projects} and \fcolorbox{black}{colorPeerR}{\ssmall \color{white}student-driven feedback}. The \rule[2pt]{10pt}{1pt} solid edges are directly utilized feedback in future projects, while \rule[2pt]{3pt}{1pt} \rule[2pt]{3pt}{1pt} \rule[2pt]{3pt}{1pt} dashed edges are indirect utilization. 
    Projects are in 4 broad categories of objectives, including: \setlength{\fboxrule}{2pt}\fcolorbox{colorFamil}{white}{\ssmall familiarization}; 
    \fcolorbox{colorFound}{white}{\ssmall foundation building}; 
    \fcolorbox{colorSkill}{white}{\ssmall applying skills in new contexts}; and
    \fcolorbox{colorSyste}{white}{\ssmall software engineering}.
    }    
    \label{fig:project_map}
    
\end{figure}

\setlength{\fboxrule}{1pt}
\setlength\fboxsep{3pt}

\section{Peer Review}
\label{sec.rubric}

Upon the completion of each project, students were asked to provide reviews of 3 randomly selected peers' work within 5-7 days. One point of variation between the semesters was the platform used for code submission and subsequent peer review. For project submissions, we experimented with zip files via Canvas assignments~(2017), Bitbucket\footnote{Bitbucket: \url{https://bitbucket.org/}}~(2018), and GitHub Classroom\footnote{GitHub Classroom: \url{https://classroom.github.com}}~(2019).

Peer review in 2017 was handled via Canvas' built-in peer review platform, which provides a liberal arts style of peer review, where a document is displayed, and questions appear alongside. In our case, students had to download and run code on their local machine. Once feedback was submitted, it would immediately become available to the recipient. In 2018, peer reviews were assigned and submitted via a Canvas quiz. At the end of the peer review period for a project, the feedback was returned by e-mail using custom scripts. In 2019, Google Forms was used to capturing feedback and delivered via a custom webpage at the end of the peer review period.

\begin{figure}[!b]
    \centering
    
    {\begin{minipage}[b]{0.32\linewidth}
        \centering
        \includegraphics[height=1.25in]{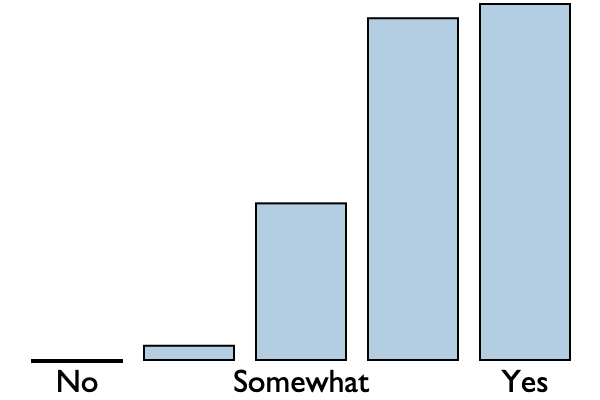} \placeImageLabelF{n=61}
        \vspace{-5pt}
        \subcaption[scriptsize]{Canvas Peer Review\label{fig:rubric:2017}}
    \end{minipage}} 
    \hfill
    {\begin{minipage}[b]{0.32\linewidth}
        \centering
        \includegraphics[height=1.25in]{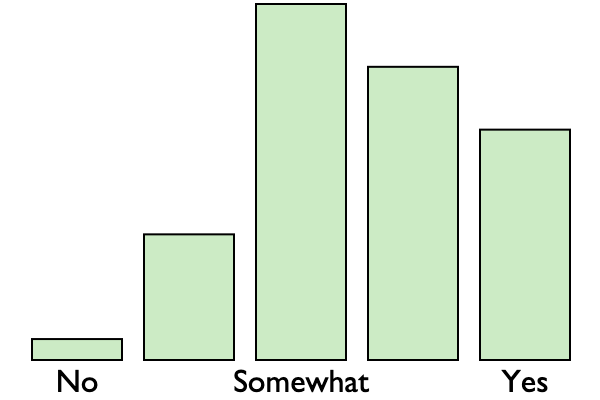} \placeImageLabelF{n=49}
        \vspace{-5pt}
        \subcaption[scriptsize]{Canvas Assignments\label{fig:rubric:2018}}
    \end{minipage}}     
    \hfill
    {\begin{minipage}[b]{0.32\linewidth}
        \centering
        \includegraphics[height=1.25in]{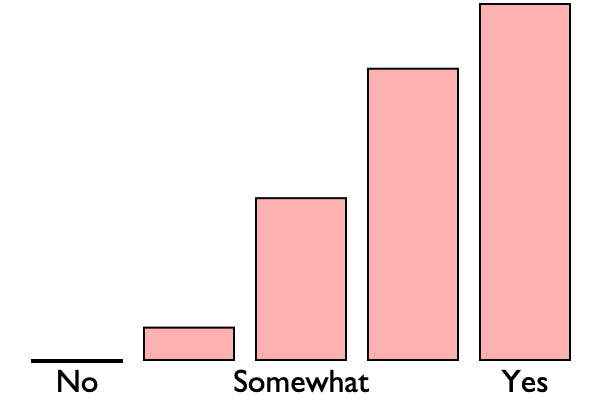} \placeImageLabelF{n=26}
        \vspace{-5pt}
        \subcaption[scriptsize]{Google Forms\label{fig:rubric:2019}}
    \end{minipage}}     

    \caption{Post-course survey results for ``how well'' the peer review interfaces worked.  \ColorCode}
    \label{fig:survey:interfaces}
\end{figure}

\autoref{fig:survey:interfaces} shows the post-course survey response to the question of whether the interface ``worked well''. These are independent samples, with no common baseline, making direct comparison impossible. However, the students using the Canvas peer review system had the most favorable view of that platform (see \autoref{fig:rubric:2017}), followed by the Goggle Forms group (see \autoref{fig:rubric:2019}), then the Canvas Quiz group (see \autoref{fig:rubric:2018}). In the free-response section of the survey, several students in the Google Forms group stated that better integration with Canvas would have improved the experience.

\subsection{Peer Review Rubric}

A rubric is a ubiquitous pedagogical tool that articulates the expectations for an assignment by listing the criteria and describing levels of quality from excellent to poor. They are used by a large number of instructors in a variety of disciplines to provide feedback on and to grade an array of student products, e.g., writings, oral presentations, portfolios, projects, etc.~\cite{li2010assessor, walker2017let, friedman2015mapping}.

\paragraph{A Rubric for Visualization Education}
We addressed the limited availability of rubrics for peer review in the visualization classroom with our rubric at the Pedagogy of Visualization 2017 workshop~\cite{Friedman2017}. We have included a copy of the rubric in our supplementary materials, and a \LaTeX version of the rubric can be downloaded/cloned/forked at \url{https://github.com/USFDataVisualization/VisPeerReview}. We briefly revisit the structure of the rubric.

The rubric was built by carefully reviewing the course content and extracting key concepts necessary for demonstrating proficiency in learning objectives \textbf{\refLTwo} and \textbf{\refLThree}. The basic structure of the rubric divides topics into 5 major assessment categories, with each category having 3 sub-assessments affixed to a 5-point scale. Each sub-assessment contains a comment box for details on the scoring.

The 5 major assessment categories are: algorithm design, interaction design, visual design, design consideration, and visualization narrative. The algorithm design category is concerned with algorithm selection and implementation. Interaction design is concerned with how the user interacts with the visualization. Visual design is concerned with the technical aspects of how data are placed in the visualization (e.g., visual encoding channels, their expressiveness, and their effectiveness). Design consideration focuses on the composition and aesthetic aspects of the visualization, such as embellishments. The final category, visualization narrative, is used in projects where the story surrounding the visualization is as important as the visualization itself.

\paragraph{Rubric Customization}
In the original design of the rubric, we intended a certain level of customization to be applied based upon the content of an assignment or course. For each project, we extracted the relevant components from the full template. Project~1, for example, included a narrative component, while no other project included such a requirement. Projects 4-8 had interaction components, while Projects 1-3 did not. Furthermore, the sub-assessments included in the early project rubrics reflected topics that had been covered in class. The rubrics for all projects can be found in the supplementary materials. Furthermore, an example can be seen in \autoref{fig:teaser}~(right) for the assessment received from 4 peers to the submission in \autoref{fig:teaser}~(left). The example shows 3 main categories, visual design, design consideration, and interaction, along with 16~sub-assessment questions for Project 6.

\begin{figure}[!t]
    \centering
    
    {\begin{minipage}[b]{0.32\linewidth}
        \centering
        \includegraphics[height=1.25in]{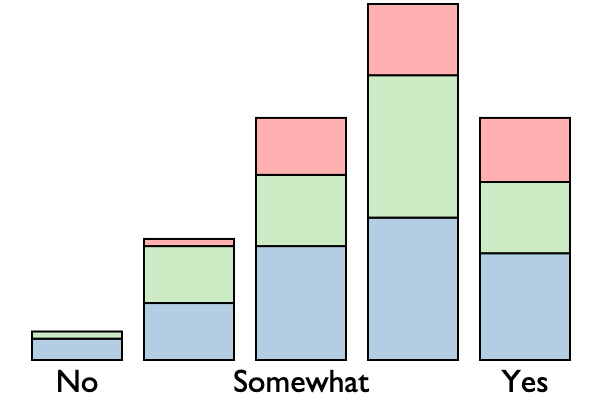} \placeImageLabelB{n=139}
        \vspace{-5pt}
        \subcaption[scriptsize]{Questions Useful\label{fig:rubric:questions}}
    \end{minipage}} 
    \hfill
    {\begin{minipage}[b]{0.32\linewidth}
        \centering
        \includegraphics[height=1.25in]{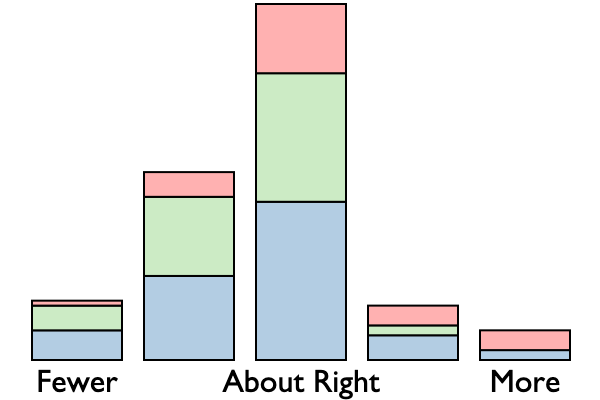} \placeImageLabelB{n=139}
        \vspace{-5pt}
        \subcaption[scriptsize]{Number of Categories\label{fig:rubric:number}}
    \end{minipage}} 
    \hfill
    {\begin{minipage}[b]{0.32\linewidth}
        \centering
        \includegraphics[height=1.25in]{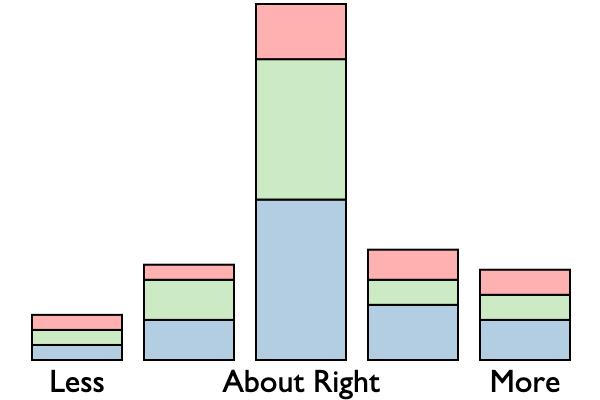} \placeImageLabelB{n=139}
        \vspace{-5pt}
        \subcaption[scriptsize]{Amount of Detail\label{fig:rubric:detail}}
    \end{minipage}} 
    
    \caption{Questionnaire results for questions about the rubric. \ColorCode}
    \label{fig:rubricQuestions}
\end{figure}

\paragraph{Rubric Evaluation}
As instructors, the rubric covers all of the key concepts we intend for students to master. On the post-course questionnaire, we asked the students 3 questions related to their opinions of the rubric (see \autoref{fig:rubricQuestions}). Their opinions indicated that, while the rubric questions were useful (see \autoref{fig:rubric:questions}), fewer (see \autoref{fig:rubric:number}) with the same level of detail  (see \autoref{fig:rubric:detail}) would be preferred. We are currently considering methodologies for reducing the number of questions, e.g., combining similar sub-assessments over the course of the semester.

\section{Peer Review and Student Learning Outcomes}

While assessment is the heart of formal higher education and a core component of effective learning~\cite{national2000people}, for this work, we do not prioritize evaluating the correctness of a visualization, i.e., how it corresponds to a professor's grade. Rather, we are interested in evaluating the \textit{influence} peer review had on students producing the visualizations.

\vspace{5pt}
\noindent
Specifically, we limit our analysis to 3 questions:

\begin{enumerate}[noitemsep,itemsep=5pt]
    \item Does peer review reinforce course content?
    \item Do students engage in and enjoy the peer review process?
    \item What aspect of peer review is most beneficial to students?
\end{enumerate}

\subsection{Does Peer Review Reinforce Course Content?}
\label{sec.contents}

We primarily utilized the student peer review comments to determine whether peer review reinforces course content. If students mention key concepts learned in the course in written responses, we interpret it to mean that they took the opportunity to identify course content in context (i.e., learning objective~\textbf{\refLTwo}).

The open-ended review comments (peer=3104, self=339, total=3443) were gathered from multiple comment sections on each review form. Each comment section was concatenated into a single string and analyzed by the process described in \autoref{sec.course.data}.

\begin{wrapfigure}[22]{r}{0.6\textwidth}
    \centering
    \vspace{-8pt}
    \includegraphics[trim=0 60pt 250pt 0, clip, width=0.95\linewidth]{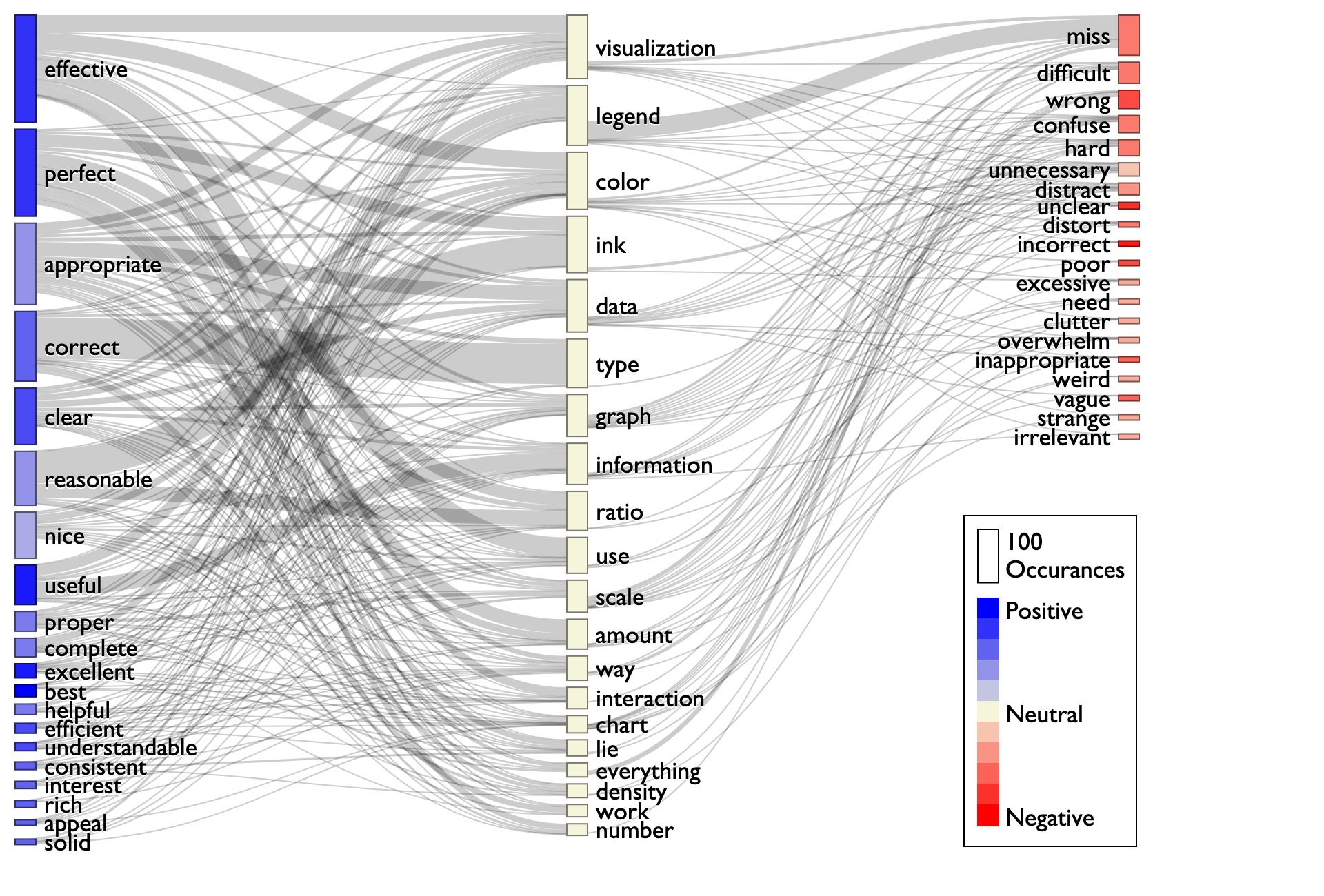}
    \caption{Visualization of the sentiment of peer reviews. Aspects (i.e., nouns) appear in the center column connected to positive (left) and negative (right) sentiment words that they appear by in text. The height of the bars indicates the number of occurrences. Color saturation indicates how positive or negative the sentiment is.}
    \label{fig:sentiment}
\end{wrapfigure}

We first compared whether students were using terminology from the rubric or whether they were commenting on other things. 
Included in the top 20 aspects are the following words: 
visualization, 
legend,
color,  
ink, 
data,
type (of data),  
graph, 
information, 
ratio (of data), 
use (of color, encoding), 
scale, 
amount (of ink, data), 
interaction, 
chart,  
lie, and 
density. 
These words are all found on the rubric, so students are likely parroting the terminology. Nevertheless, we posit that this forces the students to identify course content in the context of their peers' work. \autoref{fig:sentiment} provides some context for the utilization by displaying the aspects in the center column with the number of connected positive sentiment words (left column) and the number of connected negative sentiment words (right column).

Although we cannot directly measure whether students understand the concepts better through peer review, we can be confident that this approach provides repeated exposure to key concepts (giving and receiving on-topic feedback), and students are, at the very least, repeating terminology.

Finally, comments from the post-course questionnaire like the following seem to confirm that students note course content in each others' visualizations:

\studentQuote{I'd definitely recommend keeping it \ldots\ It also helps those struggling with concepts to see how others did it, to do it better on future assignments.}

\studentQuote{Peer review helps in understanding data visualization principles \ldots\ This helps a lot in doing future assignments and understanding my mistakes.}

\subsection{Do Students Engage In \& Enjoy Peer Review?}

To quantitatively evaluate student engagement, we analyze the number and variety of words written as recommended by~\cite{van2010effective}. We specifically tag nouns (aspects), adjectives (aspect modifiers), and adverbs (sentiment enhancers) from the peer review comments. There are a variety of other sentiment metrics provided by our algorithm (e.g., sentiment score, percent of reviews that lacked sufficient information to score, the average purity of positive sentiment) that make little sense unless comparing between individual projects. The relevant summary statistics are shown in \autoref{table:pc}.

\begin{table}[!ht]
    \centering 
    \caption{Analytic evaluation of peer comments}
    \label{table:pc}
    \vspace{-5pt}
    \renewcommand{\arraystretch}{1.2}
        \begin{minipage}[m]{9.75cm}
        \begin{tabular}{lcccc}
         & \multirow{2}{*}{Reviews} & Avg & Avg Positive & Avg Negative \\
         &                          & Keywords & Keywords & Keywords \\
         \cline{2-5}
        Graduate & 1543 & 7.75 & 4.99 & 2.76 \\
        Undergraduate & 1573 & 9.26 & 5.78 & 3.48 
        \end{tabular}
        
        \vspace{6pt}
        \begin{tabular}{rccccc}
         & Avg Words  & Avg Words    & Avg        & Avg     & Avg   \\
         & Per Review & Per Sentence & Adjectives & Adverbs & Nouns \\
        \cline{2-6}
        \hspace{15pt}G & 100.31 & 7.85 & 9.78 & 4.91 & 30.50 \\
        \hspace{15pt}U & 135.61 & 10.45 & 12.42 & 8.37 & 36.86
        \end{tabular}
        \end{minipage}
    \renewcommand{\arraystretch}{1.0}
    \vspace{-5pt}
\end{table}

We noticed that undergraduates wrote more than graduate students, with more words per sentence and a greater variety of tagged parts of speech (especially adverbs). Interestingly, the ratio of negative to total keywords was similar for undergraduates~($38\%$) and graduates~($36\%$), which is an important measure of engagement because critically evaluating a visualization requires more investment than just a cursory review---it requires applying learned concepts to explain \textit{why} something is wrong (e.g., analyzing for ``lie factor''). Considering the length, variety of parts of speech, and the ratio of negative keywords may indicate that undergraduate students are slightly more invested in the peer review process than their graduate peers. Many post-course questionnaire comments reflected a general sense of increased engagement and motivation:

\studentQuote{Process is interactive and healthy.}

\studentQuote{I enjoyed it! Saw some really good work by my peers that motivated me in the final projects.}

\studentQuote{Definitely worth the time and energy, learned a lot by examining code, which helped to see their thought process.}
\vspace{10pt}

\begin{figure}[!b]
    \centering
    
    {\begin{minipage}[b]{0.33\linewidth}
        \centering
        \includegraphics[height=1.25in]{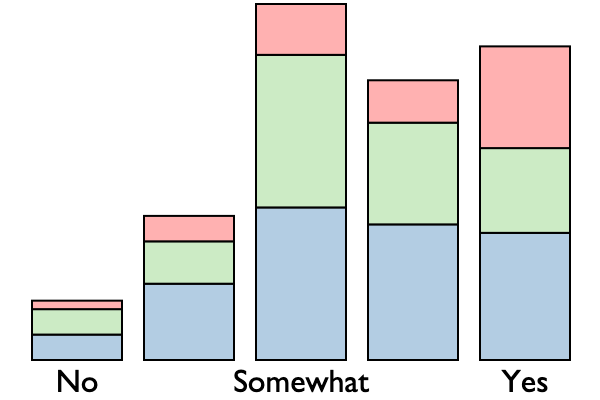} \placeImageLabelB{n=136}
        \vspace{-5pt}
        \subcaption[scriptsize]{Learned More Because of Peer Review}
    \end{minipage}} 
    \hfill
    {\begin{minipage}[b]{0.33\linewidth}
        \centering
        \includegraphics[height=1.25in]{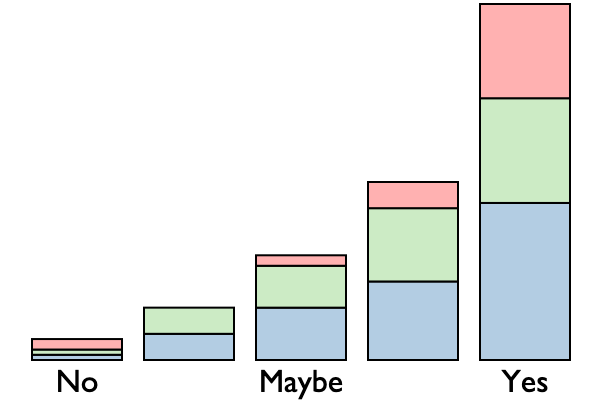} \placeImageLabelB{n=136}
        \vspace{-5pt}
        \subcaption[scriptsize]{Recommend Continuing Peer Review}
    \end{minipage}} 
    \hfill
    {\begin{minipage}[b]{0.26\linewidth}
        \centering
        \includegraphics[height=1.25in]{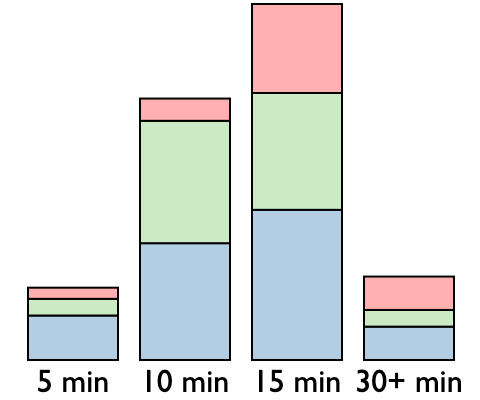} \placeImageLabelC{n=139}
        \vspace{-5pt}
        \subcaption[scriptsize]{Time Spent Per Review\label{fig:rubric:time}}
    \end{minipage}}  
    
    \caption{Questionnaire results related to (a-b) engagement and (c)~effort. \ColorCode}
    \label{fig:survey:engagement}
\end{figure}

To discover whether students enjoyed the peer review process, we asked 2 questions on the post-course questionnaires (see \autoref{fig:survey:engagement}): (1) if students believe they learned more because of the peer review process (perceived improvement); and (2) if students recommend continuing peer review. 82\% of respondents reported learning at least somewhat more (score of 3 or more; mean~=~3.6). To validate consistency across semesters, we ran unpaired t-tests comparing 2017/2018: \ttest{107}{0.32}{0.75} and 2018/2019: \ttest{73}{1.6}{0.11}, which showed no statistical difference (i.e., $p>0.5$).

Additionally, 75\% recommended continuing the process (score of 4 or more; mean~=~4.1, 2017/2018: \ttest{107}{0.65}{0.51} and 2018/2019: \ttest{73}{1.5}{0.15}) with almost half of students strongly recommending it, despite the fact that reviews were more work, taking between 10 and 15 minutes to fill out (see \autoref{fig:rubric:time}). Finally, some students enjoyed the process so much that they recommended increasing the stakes of peer review in the course:

\studentQuote{I liked being able to view what others did because it helped me see ways I can improve my work. I also think the peer reviews should be worth more points.}

\studentQuote{I think you should showcase the best and worst visualizations in class. :D Let the students praise/rip apart them as an exercise. It’s also more motivation to do well.}

\subsection{What Aspect Is Most Beneficial to Students?}
\label{sec.benefits}

\begin{figure}[!b]
    \centering

    {\begin{minipage}[b]{0.310\linewidth}
        \centering
        \includegraphics[height=1.25in]{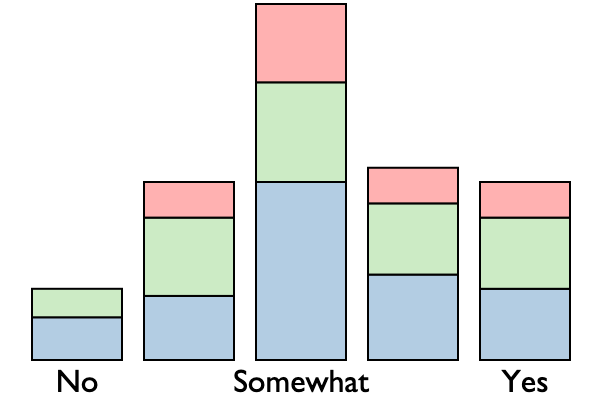} \placeImageLabelB{n=137}
        \vspace{-5pt}
        \subcaption[scriptsize]{Feedback Helpful\label{fig:survey:benefits:feedback}}
    \end{minipage}} 
    \hfill
    {\begin{minipage}[b]{0.36\linewidth}
        \centering
        \includegraphics[height=1.25in]{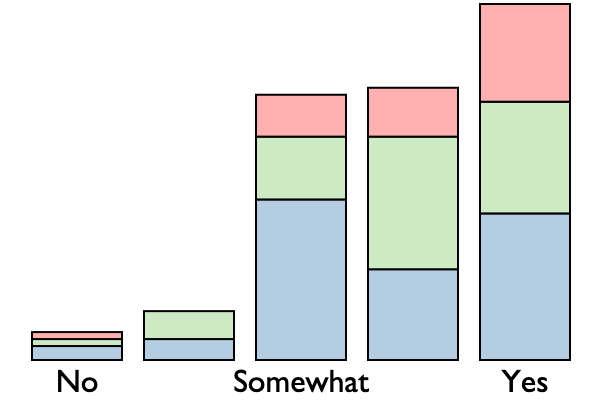} \placeImageLabelB{n=139}
        \vspace{-5pt}
        \subcaption[scriptsize]{Seeing Others Work Helpful\label{fig:survey:benefits:others}}
    \end{minipage}} 
    \hfill    
    {\begin{minipage}[b]{0.310\linewidth}
        \centering
        \includegraphics[height=1.25in]{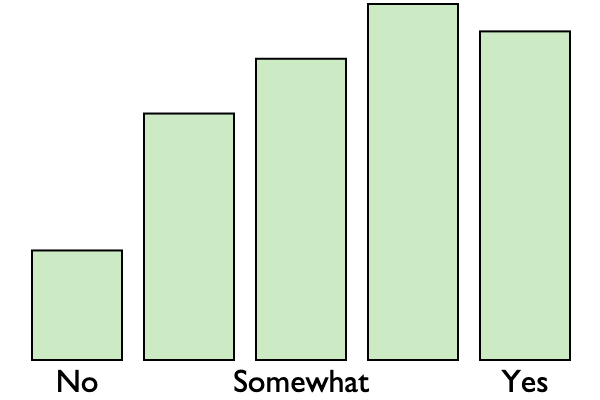} \placeImageLabelB{n=49}
        \vspace{-5pt}
        \subcaption[scriptsize]{Self-Review Helpful\label{fig:survey:benefits:self}}
    \end{minipage}}

    \vspace{2pt}
    {\begin{minipage}[b]{0.315\linewidth}
        \centering
        \includegraphics[height=1.25in]{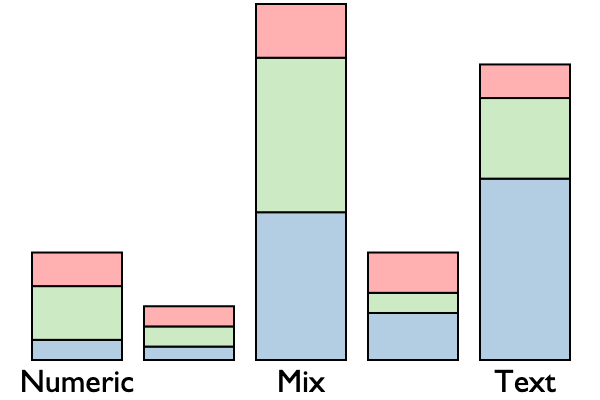} \placeImageLabelB{n=137}
        \vspace{-5pt}
        \subcaption[scriptsize]{Prefer Feedback Type\label{fig:rubric:numeric}}
    \end{minipage}} 
    \hfill    
    {\begin{minipage}[b]{0.315\linewidth}
        \centering
        \includegraphics[height=1.25in]{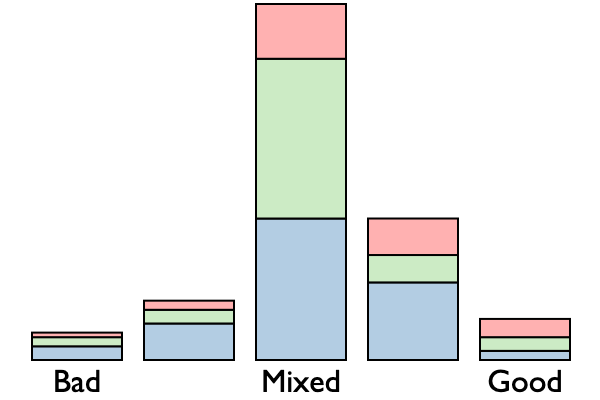} \placeImageLabelB{n=137}
        \vspace{-5pt}
        \subcaption[scriptsize]{Review Quality\label{fig:survey:benefits:quality}}
    \end{minipage}}        
    \hfill
    {\begin{minipage}[b]{0.315\linewidth}
        \centering
        \includegraphics[height=1.25in]{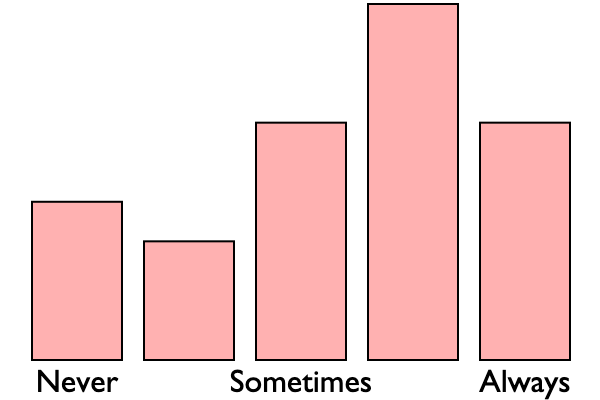} \placeImageLabelB{n=28}
        \vspace{-5pt}
        \subcaption[scriptsize]{Looked at Feedback\label{fig:rubric:usedFeedback}}
    \end{minipage}} 
    
    \caption{Questionnaire results related to the perceived utility of feedback. \ColorCode}
    \label{fig:survey:benefits}
\end{figure}

To determine which aspect of peer review is most beneficial to students, we asked 3 questions about the helpfulness of content: receiving feedback (\autoref{fig:survey:benefits:feedback}), providing feedback to others (\autoref{fig:survey:benefits:others}), and participating in self-review (\autoref{fig:survey:benefits:self}). Interestingly, in order from most helpful to least, most students found (1)~reviewing others' work (mean = 3.9, 2017/2018: \ttest{109}{0.71}{0.42} and 2018/2019: \ttest{75}{1.1}{0.28}); then (2)~self-review (mean = 3.4); and finally (3)~feedback received (mean = 3.2, 2017/2018: \ttest{109}{0.20}{0.74} and 2018/2019: \ttest{75}{0.56}{0.58}) helpful. Thus, it appears that students perceive the maximum benefit from their ability to review others' work (learning objectives \textbf{\refLTwo} and \textbf{\refLThree}), not by receiving feedback on their own, which is consistent with Garousi's findings~\cite{garousi2010applying} (see \autoref{sec:background:edu}). 
Students leaned towards preferring textual (as opposed to numeric) responses (see \autoref{fig:rubric:numeric}), but a surprisingly low rating was received for the quality of peer reviews (mean = 3.2, 2017/2018: \ttest{108}{0.08}{0.99} and 2018/2019: \ttest{74}{1.7}{0.10}) (\autoref{fig:survey:benefits:quality}) which may explain why nearly $25\%$ of students never or rarely looked at their feedback (\autoref{fig:rubric:usedFeedback}). 

Finally, many students mentioned explicitly in the course questionnaire how helpful it was to view others' code and visualizations:

\studentQuote{Looking at other people’s code helped. Grad students = bad code.}

\studentQuote{I think it's a good idea; it helped me by giving examples of what not to do mostly.}

\studentQuote{I am very impressed with this process. It is very helpful to students.}

\studentQuote{The way it’s done gives you a chance to learn from many different peers as well as help teach many.}

\begin{figure*}[!t]
	\centering
	\hfill
	\includegraphics[trim = 0 0 0 0, clip,width=0.97\linewidth]{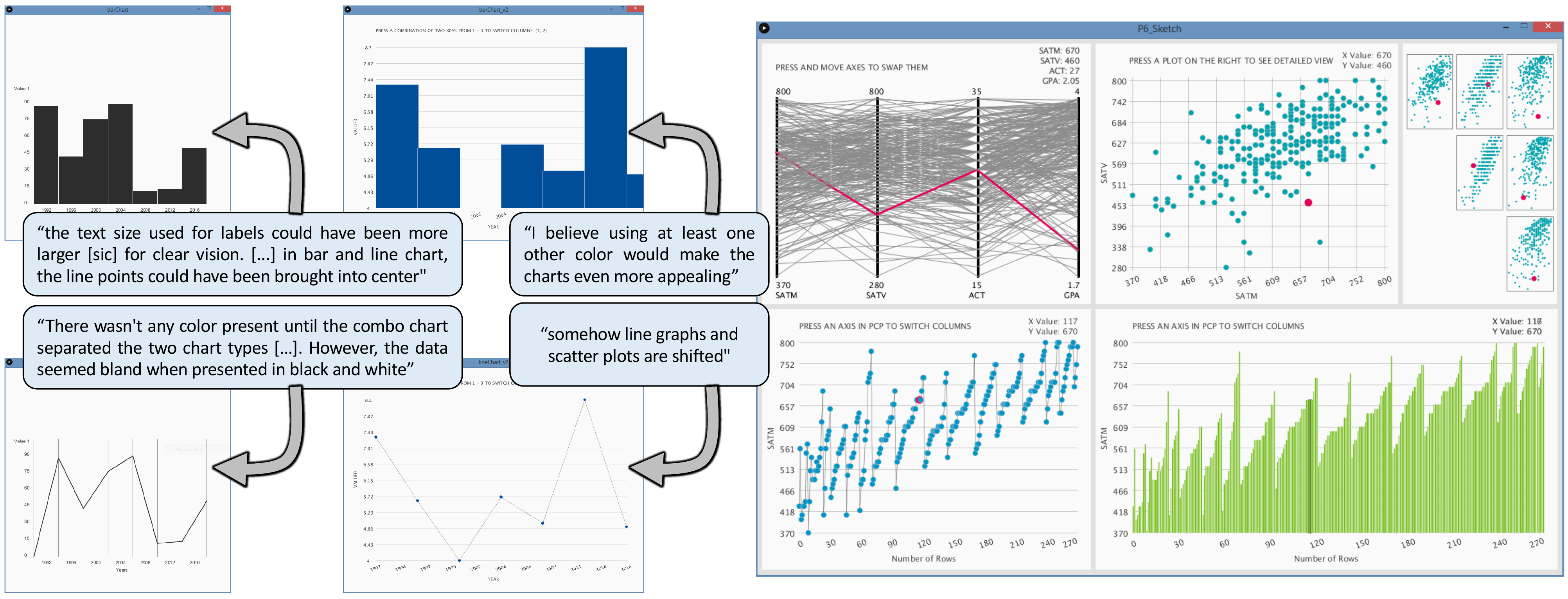}
	
	\begin{minipage}{0.225\linewidth}
    \subcaption[scriptsize]{Bar \& Line Chart from Project 2 ``Familiarization with Processing''\label{fig:proj2}}
    \end{minipage}
	\hspace{2pt}
	\begin{minipage}{0.2\linewidth}
	\subcaption[scriptsize]{Bar \& Line Chart from Project 4 ``Adding Interaction''\label{fig:proj4}}
    \end{minipage}
    \hfill
	\begin{minipage}{0.525\linewidth}
	\subcaption[scriptsize]{Dashboard from Project 6 ``Building a Dashboard''\label{fig:proj6}}
	\end{minipage}
	
	\vspace{-5pt}
	\caption{Example series of 3 projects and associated feedback utilized to improve the design.}
	\label{fig:studentFeedback}
\end{figure*}

\subsection{Peer Review Examples}

Although we did not control for peer review, the following examples of particular students' progress throughout the semester (learning objective \textbf{\refLOne}) reflect a combination of instruction and the effectiveness of the peer review process.

In the first example, we highlight differences between Projects~2, 4, and 6 to demonstrate the effect of a student receiving and implementing peer feedback. \autoref{fig:proj2} received the following comment(s): ``the text size used for labels could have been more larger [sic] for clear vision. [...] in bar and line chart, the line points could have been brought into center'' and ``There wasn't any color present until the combo chart separated the two chart types with a red color. However, the data seemed bland when presented in black and white.'' In response, the student changed the color of the bar chart, moved the line chart points to the center of the label (actually displaying the points themselves) and made the axis titles slightly larger, as shown in \autoref{fig:proj4}. \autoref{fig:proj4} received the following comment(s): ``somehow line graphs and scatter plots are shifted'' and ``I believe using at least one other color would make the charts even more appealing''. In response, the student added another color to \autoref{fig:proj6} and shifted the appropriate graphs to not overlap. Thus, in addition to benefiting from reviewing others' work, the student appeared to consider and implement much of the feedback they received.

\begin{figure*}[!t]
    \centering
    \includegraphics[width=1.0\linewidth]{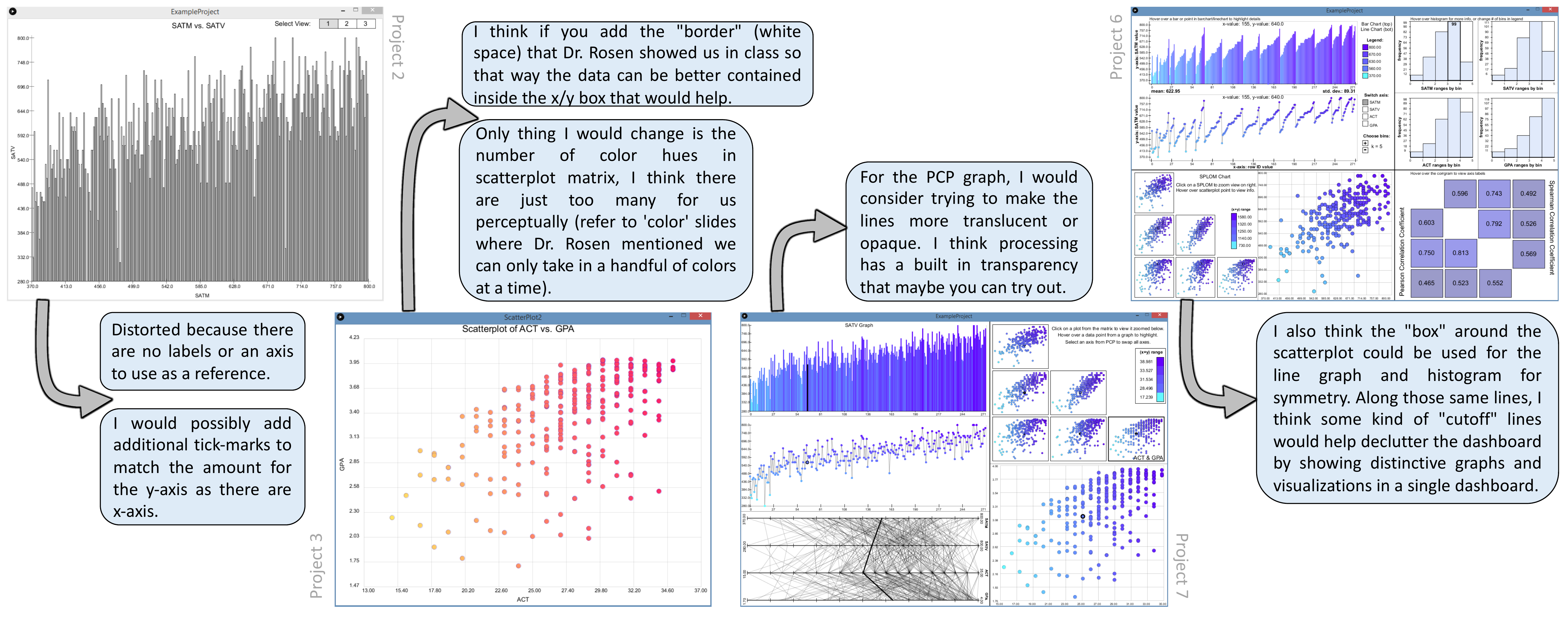}
    \caption{Illustrative example of a student's projects and the feedback they \textit{\underline{gave}} to their peers, reflecting applied concepts.}
    \label{fig:nofeedback}
\end{figure*}

In some cases, even if a student received no useful feedback, we still noticed continuous improvement, as in \autoref{fig:nofeedback}. Clearly, something was influencing the improvement, and one could speculate that seeing others' work contributed to this. However, we have no way to confirm this hypothesis.

Another interesting observation from reviewing feedback was that students tend to comment on others' work in avenues that they have already implemented in their visualizations (showing progress in learning objectives \textbf{\refLTwo} and \textbf{\refLThree}). \autoref{fig:nofeedback} shows student projects and the feedback the student \underline{\textit{gave}} to their peers. For example, in Project 2, the student mentions axis ticks and labels---something they carefully implemented in their bar and line charts. In the Project~3 feedback, the student points out the correct use of colors and directly references materials learned in class. The student references a specific programming technique for avoiding clutter in a PCP in their Project 6 feedback, and finally, a tip to delineate charts on the dashboard for Project 7 feedback. In each situation, the student offers advice that corresponds to a technique they correctly implemented, which corroborates our findings in \autoref{sec.contents} that peer review reinforces course content by allowing students to communicate recently learned and applied concepts during the peer review process (an opportunity they might not otherwise have).

\subsection{Instructor Perspective: Observed Student Benefits}

The previous examples have shown, from student data, how peer review benefits visualization students. We have observed several additional student benefits that we informally evaluate.

\paragraph{Formalizing Existing Collaborations} 
The teaching assistant and grader for the 2018 and 2019 data visualization course pointed out one advantage of peer review related to engagement, not previously discussed. Many, but not all, students will discuss projects with one another and seek feedback without instructor intervention. He noted, however, that peer review using a rubric forces (in a positive way) interaction for some students who would not otherwise interact with and offer constructive criticism to others. To some, it offers a more comfortable environment than discussion in class. In all cases, it provides structure to the feedback. Requiring feedback in this way is important to the student (to practice critical evaluation skills), as well as to the visualization (a greater variety of perspectives contribute to increased effectiveness).

\paragraph{Increasing Ownership} 
Multiple students commented on the value of seeing each others' work. There are 2 effects that we observed: (1)~it enables seeing how peers have solved a problem in order to redesign your visualizations; and (2)~there is pressure to perform better when you know your peers will see your work. From our conversations with students, we felt that this gave the students more ownership in the projects, many discussing with pride the final product they developed.

\paragraph{Critical Evaluation Skills} 
Critical evaluation of others' work is a mandatory skill for academia, for assessing research quality, and in industry, for code and design reviews. Many engineering programs are knowledge and skills-based (\href{https://www.abet.org/accreditation/accreditation-criteria/criteria-for-accrediting-computing-programs-2018-2019}{ABET accreditation} criteria reinforce this) while teaching critical evaluation informally. Peer review can help fill this gap in developing skills.

\section{Implementation in Other Courses}

This paper intends to encourage other educators to adopt and customize the methodology for their courses. By-and-large, the approach we have used is not restricted to the style of computer science visualization course we taught but is highly generalizable to other design-oriented or non-computer science visualization courses. Most components can be swapped out---e.g., the projects can be swapped for other projects, as long as they emphasize iterative design; the rubric can be replaced by other evaluation criteria, e.g., Design Activity Framework worksheets~\cite{mckenna2017worksheets}; etc. 

\subsection{Implementation Options}
We provide our course and recommendations as a road-map for other instructors with 4 routes to implement peer review in other classrooms:

\vspace{5pt}
\uline{Simple}: Use our project and rubric designs. This approach is less than ideal for most since each instructor has their own perspective on the most important concepts and skills to test. 
    
\vspace{5pt}
\uline{Lightweight Integration}: Peer review can be added to existing projects using our rubric or a rubric developed by the instructor. This approach lacks the feedback cycle but will still capture some of the peer review benefits, such as reinforcing course concepts. 
    
\vspace{5pt}
\uline{Middleweight Integration}: If the course already uses a continuous design process (e.g., a semester-long design project), incorporate one or more peer review stages using our rubric or a rubric developed by the instructor. In addition to reinforcing concepts, this integration enables using feedback for a continuous-improvement process.
    
\vspace{5pt}
\uline{Heavyweight Integration}: The course projects can be completely redesigned, such that they are built around peer review (instead of the other way). Projects would be specifically designed to emphasize course topics, with each project cycle building on the previous. This approach would capture all of the benefits of peer review.

\paragraph{Platforms for Delivery}
The platform for delivering content can have a significant impact on student perception of that content. Of the 3 platforms we used (see \autoref{sec.rubric}), each had advantages and disadvantages, but in the end, we found none of them ideal. Most likely, an instructor adding peer review to their course would choose the peer review mechanism already available on their LMS.

\paragraph{Instructor Effort} 
On the surface, peer review appears to potentially result in substantial time savings for instructors no longer needing to give subjective feedback. For example, 60 students, 8 assignments per semester, and 10 minutes of feedback per assignment should result in over 80 hours saved. However, much of that time is not saved but actually diverted to other activities. On one hand, additional administrative tasks arise---grades still need to be assigned, peer review requires additional effort to set up and administer, and peer reviews need to be monitored for quality. At the same time, there is an opportunity and need to provide more detailed instructor feedback to those highly engaged students who request it---we regularly advertised that instructor feedback was available, in person, upon request (e.g., during office hours). Peer review is a complement to existing educational approaches, but the instructor, as the expert, still needs to be involved.

\paragraph{Completion and Quality} 
Incentives in peer review have been widely studied in the context of writing courses~\cite{gielen2010improving, panadero2013use, van2006design, van2006peer}. We achieved a high completion rate, 95\%, by making peer reviews a mandatory part of the course, spot-checked, and graded on completion. Unfortunately, the quality of the peer reviews students received varied (see \autoref{fig:survey:benefits:quality}). Just over $56\%$ of the students rated the quality of reviews they received as ``Mixed'', with only $6\%$ stating that the quality was ``Always Good''. Increasing the stakes (i.e., making peer review quality a significant of the grade received) would likely improve quality, but create substantial overhead for instructors, as detailed analysis of every peer review submitted would be required.

\subsection{Risks}

When considering using peer review, the risk for collusion, malice, and cheating is great, even with double-blind reviews.

\paragraph{Peer Review for Grading}  
Instructors should take care to randomize peer reviewers, and grades should (at most) only be loosely based upon the results of the peer review. However, it is also important to remember that there is both art and science to design with no single optimal design, making the wisdom of the crowd potentially useful in assessment. Building machine learning models that utilize peer evaluation to (semi-)automatically assign grades is an exciting direction for future study. 

\paragraph{Collusion} 
If reviews are used for grading, the problem of collusion or malice may play a role (e.g., the review of a friend may be overly optimistic). Keeping all submissions anonymous is helpful, but anonymity is hard to maintain, especially in a smaller educational setting where students may talk and discover they are evaluating each others' visualizations.

\paragraph{Cheating} 
Providing access to other students' projects is an obvious risk. In fact, we have dealt with 2 such cases of students stealing others' code through peer review. However, frequent and repeated warnings, along with detection systems, e.g., MOSS (Measure Of Software Similarity)~\cite{schleimer2003winnowing}, can reduce the number of such incidences.

\paragraph{Falling Behind} 
Since many projects build on one another, students who fail to complete early projects to a high standard may continue to struggle on future projects in a snowball-like effect. We mitigate this by providing a significant amount of time, almost 1 month for the major integration project (Project 6). Nevertheless, the risk of frustrating and losing students early remains.

\section{Conclusion}

We have presented our experience and evaluation of peer review in the visualization classroom. This approach has 2 significant benefits. First, it provides a framework for engaging students through critical evaluation of visualizations. Second, it is a mechanism for providing students with diverse and timely feedback on their work.

Several implementation issues deserve further study. 
We utilized a variety of peer review interfaces, all of which had inflexible designs, not well suited to visualization peer review. Having a tool that enables us to describe the grammar of a graphic concisely would help to gain insights into the structure that underlie statistical and programming languages to produce better graphics~\cite{wilkinson2012grammar, wickham2010layered}.

The rubric itself has a focus on low-level details of design (e.g., visual encodings, tick-marks, and labels), as opposed to high-level design, such as composition and choice of technique. This focus is partially an artifact of our project design. We do not believe that the rubric, as presented, will be the final static version. We anticipate it will be a growing and evolving document as community members provide their input and the focus of the visualization community changes. Nevertheless, further evaluation of potential rubric designs should be considered.

A final direction is the customization of the rubric through reducing constraints. Assuming the critical thinking skills of students are weak, particularly in the domain of visualization, the rubric itself can be a tool to help improve those skills. At the beginning of the course, the rubric can include all scoring categories, and as the course progresses, categories can be combined and removed. In this way, students will go from highly structured to free-form evaluation.

\paragraph{Disclosure} 
The first author (Beasley) was a student in the 2017 graduate course. His participation in the project began \textit{after} the course was completed.

\section*{Acknowledgments}

We thank Ghulam Jilani Quadri for providing his perspective as a teaching assistant for this course. This project was supported in part by the National Science Foundation (IIS-1845204).

\bibliographystyle{abbrv}
\bibliography{main}

\end{document}